\documentclass[conference]{IEEEtran}
\usepackage{graphicx}
\usepackage{amstext}
\usepackage{algorithm}
\usepackage{algorithmic}
\usepackage{mathrsfs}
\usepackage{amssymb}
\usepackage{amsmath}
\usepackage{bm}
\allowdisplaybreaks[4]
\usepackage{epstopdf}
\usepackage{multicol}
\usepackage{stfloats}
\usepackage{url}
\usepackage{subcaption}
\usepackage{color}
\usepackage{enumerate}
\usepackage{comment}
\DeclareMathAlphabet\mathbfcal{OMS}{cmsy}{b}{n}
\usepackage{breqn}
\usepackage{bbm}
\usepackage{cite}
\usepackage{caption}
\usepackage{enumitem}
\usepackage{array}
\ifCLASSINFOpdf
\else
\fi
\newcommand\blfootnote[1]{%
  \begingroup
  \renewcommand\thefootnote{}\footnote{#1}%
  \addtocounter{footnote}{-1}%
  \endgroup
}

\begin{document}

\title{Position Aware 60 GHz mmWave Beamforming for V2V Communications Utilizing Deep Learning}

\author{\IEEEauthorblockN{Muhammad Baqer Mollah$^1$, Honggang Wang$^2$, and Hua Fang$^3$}
\IEEEauthorblockA{$^1$Dept. of Electrical and Computer Engineering, University of Massachusetts Dartmouth, MA 02747 \\ $^2$Dept. of Graduate Computer Science and Engineering, Yeshiva University, NY 10016 \\
$^3$Dept. of Computer and Information Science, University of Massachusetts Dartmouth, MA 02747 \\
Emails: mmollah@umassd.edu, honggang.wang@yu.edu, and hfang2@umassd.edu}
        }

\markboth{}%
{Shell \MakeLowercase{\textit{et al.}}: Bare Demo of IEEEtran.cls for IEEE Journals}

\maketitle

\begin{abstract}
    Beamforming techniques are considered as essential parts to compensate the severe path loss in millimeter-wave (mmWave) communications by adopting large antenna arrays and formulating narrow beams to obtain satisfactory received powers. However, performing accurate beam alignment over such narrow beams for efficient link configuration by traditional beam selection approaches, mainly relied on channel state information, typically impose significant latency and computing overheads, which is often infeasible in vehicle-to-vehicle (V2V) communications like highly dynamic scenarios. In contrast, utilizing out-of-band contextual information, such as vehicular position information, is a potential alternative to reduce such overheads. In this context, this paper presents a deep learning-based solution on utilizing the vehicular position information for predicting the optimal beams having sufficient mmWave received powers so that the best V2V line-of-sight links can be ensured proactively. After experimental evaluation of the proposed solution on real-world measured mmWave sensing and communications datasets, the results show that the solution can achieve up to 84.58\% of received power of link status on average, which confirm a promising solution for beamforming in mmWave at 60 GHz enabled V2V communications. \blfootnote{This paper has been accepted to present at 2024 IEEE International Conference on Communications (ICC), Denver, CO, USA.}
\end{abstract}


\IEEEpeerreviewmaketitle

\section{Introduction}
    \blfootnote{\copyright 2024 IEEE. Personal use of this material is permitted. Permission from IEEE must be obtained for all other uses, in any current or future media, including reprinting/republishing this material for advertising or promotional purposes, creating new collective works, for resale or redistribution to servers or lists, or reuse of any copyrighted component of this work in other works.} As a key enabler for communications technologies in connected and autonomous vehicles (CAVs) domain, the utilization of millimeter-Wave (mmWave) bands (e.g., 28 GHz and 57-71 GHz bands) has been promised to bring an abundance of spectrum resources \cite{mollah2023mmwave, ciaramitaro2023signalling}. For example, in the context of cooperative perceptions like high-throughput and low latency demanding applications in CAVs, the connected vehicles mainly exchange a vast amount of 3D point cloud data from LiDAR sensors instead of lightweight textual data \cite{ngo2023cooperative}. Realizing the high throughput and low latency benefits, mmWave communications enabled by vehicle-to-vehicle (V2V) communications pave the way towards accessible and safe autonomous transportation systems.
    
    On the other side, one inherent property of mmWave band is having high path attenuation of its signals, leading to drastically impact on the link quality performance degradation \cite{sun2018propagation}. In particular, beamforming techniques are utilized to address this bottleneck, whereby massive antenna arrays are typically employed including formulating the narrow beams, thereby achieving desired high throughput from mmWave communications. Likewise, the narrow beams are expected to be aligned precisely and even required to re-direct in accordance with the environmental settings and any changes. Typically, in codebook based beamforming, the beam selection techniques are applied to find the optimal beams over a number of pre-defined beam codebooks.

    In practice, with standard defined beam selection techniques, the vehicles usually select the best beam pairs through an exhaustive beam measurements process, which basically introduce computing and latency overheads while applying in highly dynamic moving vehicles. This overhead problem occurs mainly due to tight contact times (the time period of receiver's received the correct packets), frequent beam realignments, and change of channel state estimation to perform beam computing. Hence, research efforts are crucial to avoid beam misalignment along with reduced and reasonable overheads in order to take the fully potential benefits from mmWave communications. 
    
    Considering this challenge in vehicular settings, recent works have suggested effective approaches to configure the communications links through leveraging the out-of-band contextual information, which can be obtained from other lower bands (sub-6 GHz bands) \cite{kyosti2023feasibility, alrabeiah2020deep}, RADAR communications bands \cite{graff2023deep, luo2023millimeter, demirhan2022radar}, or extracted useful sensing information from on-board vehicular non-RF sensing devices \cite{gu2023meta, xu2023computer, gutune2023}. For instance, deployment based the 3GPP 5G NR (New Radio) \cite{lin2022overview} and IEEE 802.11ad \cite{misc2} standards might take around 10 milliseconds to select the best beams through searching over all beam directions and need to run over repeatedly when the vehicles move forward. In this regard, combining the surrounding perception capable side-information collected from the vehicle sensors and then, exchanging with sub-6 GHz channels has recently been demonstrated in work \cite{salehi2022flash} as promising results in terms of decreasing the end-to-end latency, which helps to reduce 52.75\% on average time in IEEE 802.11ad standard. Besides, another work in \cite{salehi2022deep} has observed regarding the beam search space by utilizing almost similar side-information, 80\% and 50\% on average overhead caused by beam selection can be reduced for 3GPP 5G NR and IEEE 802.11ad, respectively.

    Over the recent years, several approaches pursuing efficient beamforming have been introduced based on the out-of-band position specific information. In particular, the solution in \cite{morais2023position} has leveraged machine learning technique to investigate how position information can help to reduce the beam training overheads. Likewise, the work in \cite{rezaie2022deep} emphasizes on both location and orientation information to perform 3D beam selection by utilizing deep learning technique. Further, the authors in \cite{va2017inverse} have proposed an approach, where a multi-path fingerprint-based database is maintained which records fingerprints of necessary location information along with channel characteristics so that the knowledge of reliable beamforming can be known in advance. Besides, almost similar concepts have been utilized in \cite{mattos2022geolocation} and \cite{khosravi2022location}. In \cite{mattos2022geolocation}, the authors utilize spatially indexed historical information so that statistically significant beam sectors can be chosen in specific locations. This proposed work is particularly focused on IEEE 802.11ad standard with an aim of reducing the time of beam sweeping, thereby improving the available time for data communications. And, the work in \cite{khosravi2022location} has utilized the information about spatial correlation of stronger channels in-between the receiver and sender at given locations so that the users are able to track the information beforehand to establish communication link quickly.
    
    However, the aforementioned works are mainly limited to either for vehicle-to-infrastructure (V2I) communications or based on ray-tracing simulations, but utilizing real-world measured wireless data as well as considering the mobility of connected vehicles under V2V connectivity have not been investigated yet. With these limitations in mind, we introduce a solution for mmWave beamforming for V2V communications with the aid of a proposed deep learning model. Specifically, the deep learning model takes vehicular position information as input and predicts a subset of beams, that is, top-$M$ beams, thus, significantly lowers down the beam search space. Different from other out-of-band enabled approaches, the proposed solution fully leverages the position data obtained from the global positioning sensors installed in the connected vehicles, subsequently, lowering down the complicacy. Finally, we validate our proposed solution with experiments on real-world 60 GHz mmWave sensing and communications datasets, compare with state-of-the-art approach, and present the results in terms of two meaningful matrices, namely accuracies and received power ratios.

\section{System Model and Problem Statement}
    In this section, we first present our considered system model, and then, we describe the problem statement.

\subsection{System Model}
\begin{figure} [h]
    \includegraphics[width=\linewidth]{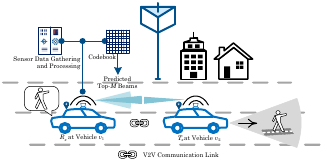}
    \centering
    \caption{Illustration of considered mmWave enabled V2V communications system model.}
    \label{fig: System-Model}
\end{figure}

    In this work, we consider a mmWave communications system model operating at 60 GHz frequency band as described in Fig. \ref{fig: System-Model}. here, the system model is comprised of a moving vehicle $v_2$ to provide communications services, e.g., cooperative perception, to another moving connected vehicle $v_1$ within its coverage area. The $v_2$ works as a transmitter $T_x$ measures its own location information, that is, latitude and longitude, and shares the information with $v_1$ in real-time manner, while the $v_1$ collects and processes the received location information and works as a receiver $R_x$. Both vehicles, $v_1$ and $v_2$, are connected with each other by vehicle-to-vehicle (V2V) communication links.

    The connected vehicle $v_1$ consists of $N_R$ antenna array elements, fixed beam codebooks $\mathbfcal{Q}$, and each codebook describes particular beam orientation, consequently, has received power at every beams. The beam codebooks can be expressed as $\mathbfcal{Q} = \{q_1, q_2, ..., q_{|\mathbfcal{Q}|} : q_i \in \mathbb{C}^{N_R \times 1}\}$, where $|\mathbfcal{Q}|$ denotes the total number of beamforming vectors. Besides, the antenna array elements enable $v_1$ to perform beamforming so that it can obtain adequate received power. Whereas, the vehicle user $v_2$ has, for the purpose of simplicity, one antenna and assuming it is oriented always towards the $v_1$. Thus, the total beamforming vector can be represented by $(\mathbfcal{Q} \times 1)$. 
    
    We further consider that OFDM is used in the downlink communications with $\mathcal{N}$ subcarriers. For a given transmitted symbol $s_n \in \mathbb{C}$, the downlink received signal at $n^{th}$ subcarrier at discrete time instant $t$ can be written as:

    \begin{equation}
    x_n[t] = \mathbf{h}_n^T[t] q_i[t] s_n + \omega_n[t]
    \end{equation}

    
    where, $\mathbf{h}_n[t] \in \mathbb{C}^{N_R \times 1}$ is the channel vectors between $v_2$ and $v_1$ and $\omega_n[t]$ is the received noise.
    
\subsection{Problem Statement}
    The primary task of this work is to make optimal beams predictions from the received powers, and from the received signal $x_n[t]$, we can get the received power by summing over $\mathcal{N}$ subcarriers at the $v_1$ side for the receiver codebook $i$ as:

    \begin{equation}
    \mathcal{P}_i = \sum_{n=1}^{\mathcal{N}} |\mathbf{h}_n^T[t] q_i[t] s_n|^2
    \end{equation}

    Given the position sensing data and mmWave received power, determining the optimal beam $q_i^*$ at the $v_1$ from the beam codebook $\mathbfcal{Q}$ is referred as maximizing the received power. However, if we convert beam codebooks $\mathbfcal{Q} = \{q_1, q_2, ..., q_{|\mathbfcal{Q}|}\}$ into a unique index as $\mathcal{I} \in \{1, 2, 3, ..., \mathcal{M}: \mathcal{M} \leq |\mathbfcal{Q}|\}$, predicting optimal beam from beam codebooks can be essentially equivalent to predicting optimal beam index. Then, we can formulate predicting the optimal beam problem as:

\begin{equation}
    \mathcal{I}^* = \underset{i \in \{1, 2, 3, ..., |\mathbfcal{Q}|\}}{\text{arg max}} \mathcal{P}_i
\end{equation}
 
    In this work, our goal is to predict a set of recommended beams $\mathcal{B} = \{\mathcal{I}_1, \mathcal{I}_2, \mathcal{I}_3, ..., \mathcal{I}_M\}$ as top-$M$ beams such that $\mathcal{I}^* \in \mathcal{B}$ as top-1 beam.
    
\section{Proposed Deep Learning Model}
    In this section, we describe the proposed solution by first discussing the data preprocessing steps, followed by elaborating the proposed deep learning model architecture.

\subsection{Position Data Preprocessing}
    The raw position data captured from GPS sensors are basically in Decimal Degrees, specifically the latitude and longitude values are from $-90^{\circ}$ to $+90^{\circ}$ and $-180^{\circ}$ to $+180^{\circ}$, respectively. However, in our proposed solution as presented as next subsection, the position data will be fed into a deep learning model as input, and the input is desired to be fixed in size. In this regard, feeding raw position data directly may require developing deep learning model with high complexity in terms of architecture and computational cost. Thus, we first pass through the raw position data into a preprocessing step, namely data normalization, to make fit as well as accelerate the convergence of the proposed deep learning model. We then define the position matrix as $X_{pos}[t] \in \mathbb{R}^{N \times 2}$, where $N$ is the total number of samples, and the number $2$ is due to having two values at each time instant $t$. After that, we calculate the normalized values of latitude $x_{lt}^{\prime}$ and longitude $y_{lg}^{\prime}$ at each rows of the matrix  as: $(x_{lt}^{\prime}, y_{lg}^{\prime}) = ((x_{lt} - x_{lt\_min})/(x_{lt\_max}) - x_{lt\_min}), (x_{lg} - x_{lg\_min})/(x_{lg\_max} - x_{lg\_min}))$, within Cartesian coordinate system, where $x_{lt}$ and $x_{lg}$ are the raw latitude and longitude values, while the $max$ and $min$ denote their maximum and minimum values, respectively. Finally, the revised matrix with normalized position data is denoted as $X^{\prime}_{pos}[t]$, and $g_t$ is considered as a variable representing the vector elements of $X^{\prime}_{pos}[t]$.

\subsection{Proposed Model Architecture}

\begin{figure*} [!t]
	\includegraphics[width=.7\linewidth]{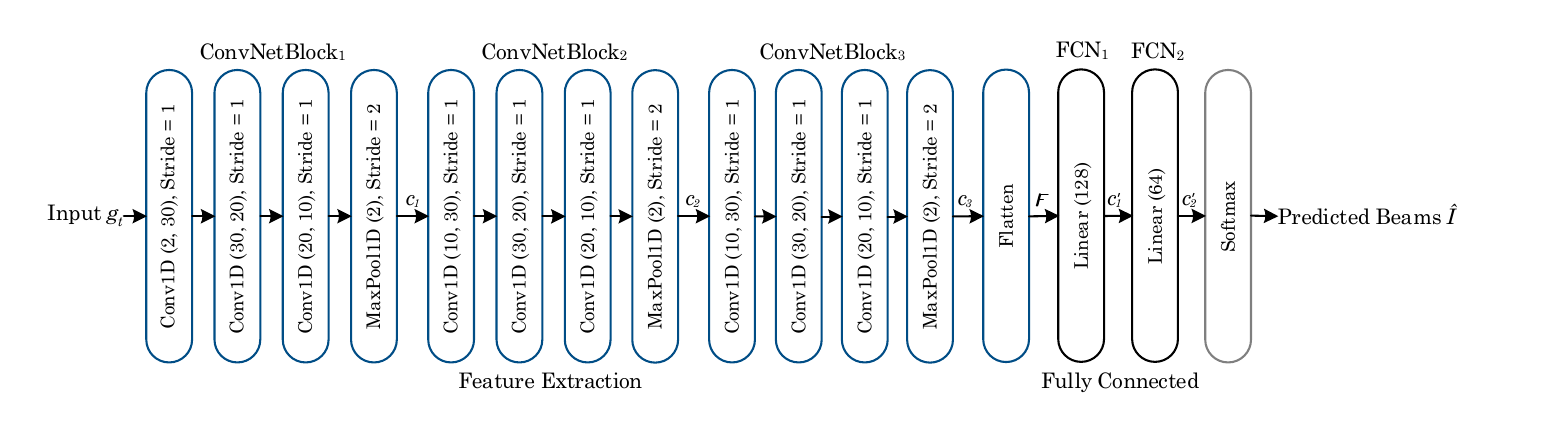} 
    \centering
    \caption{Proposed deep learning model for mmWave beamforming.}
    \label{fig: GPS-DL-Model}
\end{figure*}

    The designed architecture of proposed deep learning model for position data assisted mmWave beamforming task is presented in Fig. 2. The proposed model adopts a convolutional neural network architecture including four components: the convolutional blocks, a flatten layer, and the fully connected layers, followed by a softmax layer, and these components are elaborate as follows.
    
    • \textit{Convolutional Blocks:} The convolutional blocks, the key building blocks used in our proposed model, are mainly made up of three convolutional blocks. These convolutional blocks are formed together with three sequence of 1D (one-dimensional) convolutional layers, however, each blocks are followed by a max-pooling layer.

    • \textit{Flatten Layer:} After the convolutional blocks, one flatten layer is employed to flatten the output received from the last convolutional block, while not effecting the batch size.
    
    • \textit{Fully Connected Layers:} Two fully connected layers are incorporated to process the outputs from the flatten layer.

    • \textit{Softmax Layer:} At last, the softmax layer is involved as last layer, where softmax activation function is applied on the output from flatten layer to transform the values into probability scores within the range $[0, 1]$ with total value 1, essentially, can be interpreted to eventually determine the top-$M$ beams.

    Given the sequence vector $g_t$ as an input to the proposed model, the convolutional block $\mathsf{ConvNetBlock}_1$ first compute it to extract the features by learning the hidden meaningful representations. This block is repeated two more times as $\mathsf{ConvNetBlock}_2$ and $\mathsf{ConvNetBlock}_3$ in the network to make enhanced feature extraction and improved performance. However, the max-pooling layers in both blocks contribute to downsize the feature maps while keeping the most significant features, leading to decreasing the risk of the model becoming overfitted. Then, the output from $\mathsf{ConvNetBlock}_3$ is passed though the flatten layer. The overall process is presented as follows.

\begin{equation}
    \begin{array}{l}
    c_1 = \mathsf{ConvNetBlock}_1(g_t) \\
    c_2 = \mathsf{ConvNetBlock}_2(c_1) \\
    c_3 = \mathsf{ConvNetBlock}_2(c_2) \\
    \digamma = \mathsf{Flatten}(c_3),
    \end{array}
\end{equation}

    where, $c_1$, $c_2$, and $c_3$ are the hidden vectors and $\digamma$ represents the result from the flatten layer. This resulting output $\digamma$, after that, is forwarded to the two fully connected (dense) layers $\mathsf{FCN}_1$ and $\mathsf{FCN}_2$. The two fully connected layers together encompass a prediction by applying weights. If $c_1^{\prime}$ and $c_2^{\prime}$ denote the outputs from the two connected layers, respectively, we can describe as
    
\begin{equation}
    \begin{array}{l}
    c_1^{\prime} = \mathsf{FCN}_1(\digamma) \\
    c_2^{\prime} = \mathsf{FCN}_2(c_1^{\prime}).
    \end{array}
\end{equation}

    In the end, softmax layer ($\mathsf{Softmax}$) helps to give the final top-$M$ beam prediction result as $\hat{\mathcal{I}} = \mathsf{Softmax}(h_2^{\prime})$ by converting the unnormalized output from the connected layers into normalized values (a probability distribution). However, since our proposed model is particularly designed to predict the top-$M$ beams, the cross-entropy is utilized as a loss function with an aim to minimize the loss, which can be calculated by

\begin{equation}
    \mathcal{L}(\hat{\mathcal{I}}, \mathcal{I}) = - \frac{1}{N_b}\sum_{k=0}^{N_b-1}\sum_{l=0}^{M-1}\mathcal{I}^{(k)}\log\hat{\mathcal{I}}^{(l)},
\end{equation}

    where, $N_b$ represents the number of samples in each batches, $M$ denotes the total number of beam indices, $\mathcal{I}^{(k)}$ is the ground truth beams indices for $k^{th}$ sample, and $\hat{\mathcal{I}}^{(l)}$ is the predicted probabilities of beams indices by softmax for $l^{th}$ sample. Further, we utilize a stochastic optimization technique namely adaptive moment estimation (Adam) \cite{KingBa15} for faster convergence on minimizing the loss function while training the model.

\begin{figure*} [!t]
\begin{subfigure}[b]{\textwidth}
	\centering
	\includegraphics[width=9cm]{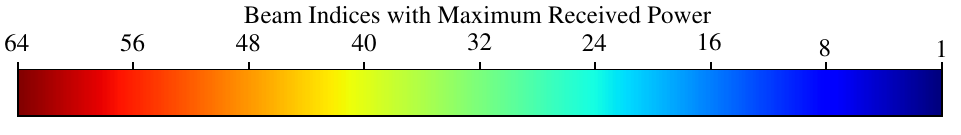}
\end{subfigure}
\begin{subfigure}[b]{0.24\textwidth}
    \centering
    \includegraphics[width=4.4cm]{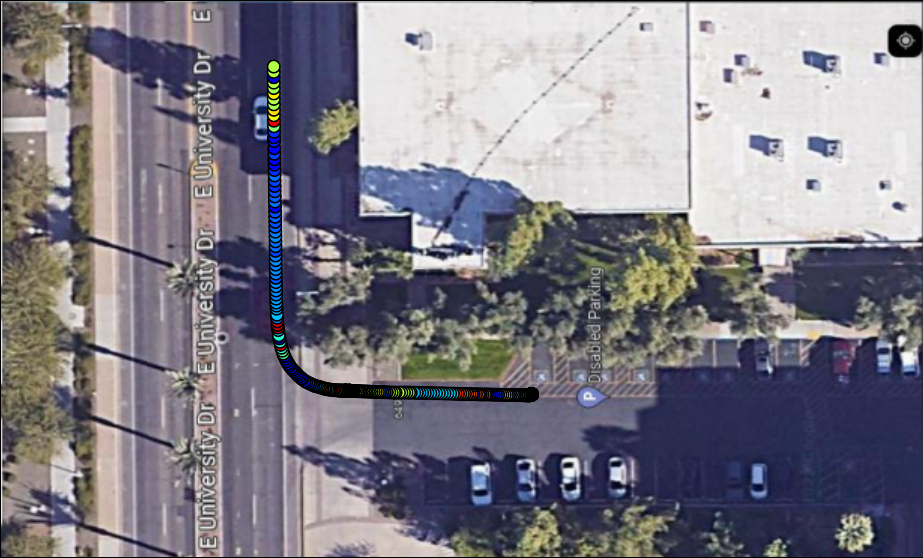}
	\caption{\centering \scriptsize For scenario 36}
\end{subfigure}
\begin{subfigure}[b]{0.24\textwidth}
	\centering
	\includegraphics[width=4.4cm]{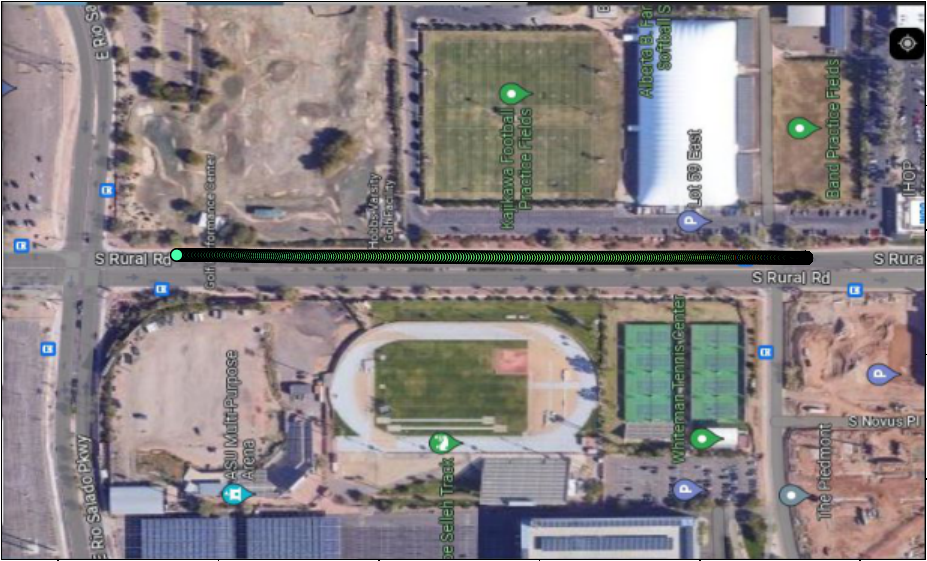}
	\caption{\centering \scriptsize For scenario 37}
\end{subfigure}
\begin{subfigure}[b]{0.24\textwidth}
    \centering
    \includegraphics[width=4.4cm]{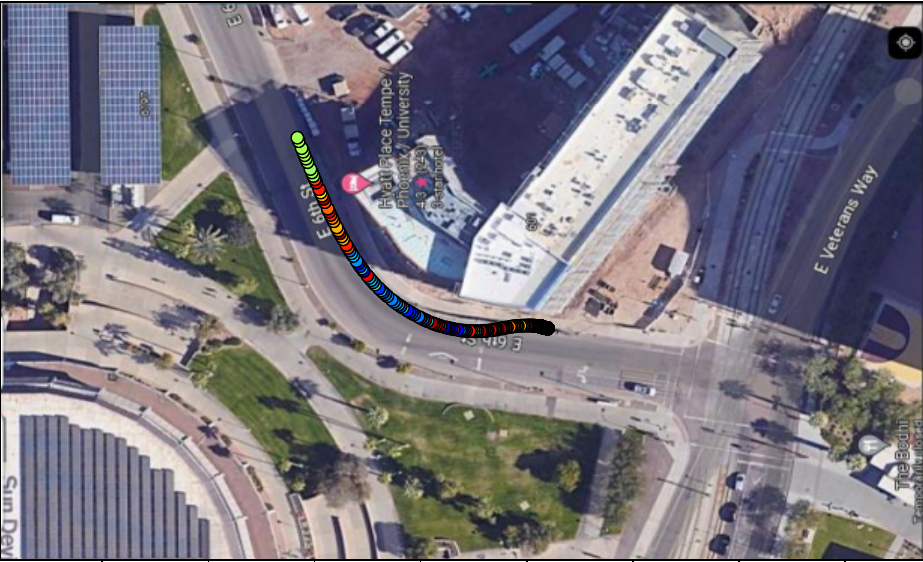}
    \caption{\centering \scriptsize For scenario 38}
\end{subfigure}
\begin{subfigure}[b]{0.24\textwidth}
    \centering
    \includegraphics[width=4.4cm]{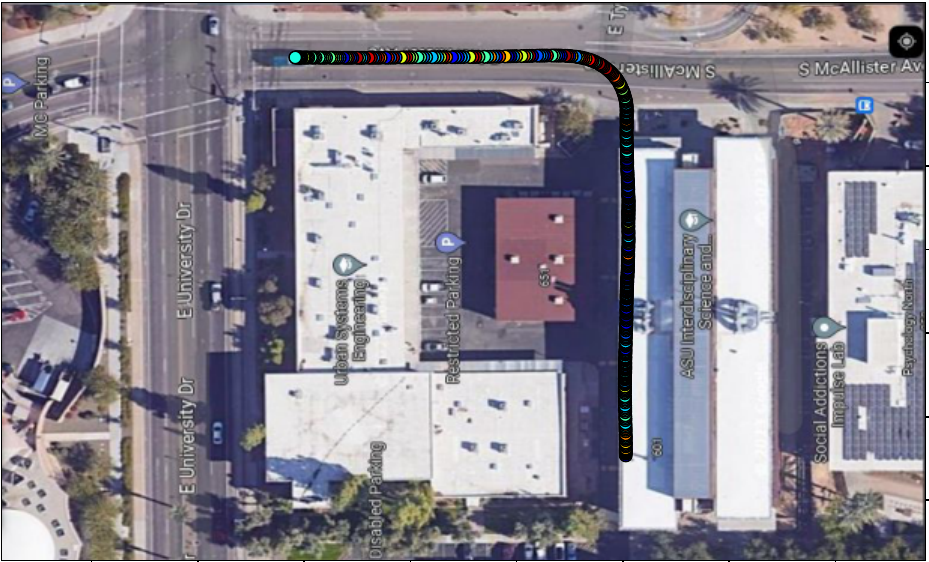}
    \caption{\centering \scriptsize For scenario 39}
\end{subfigure}
    \caption{Visual representation of receiver vehicle’s GPS location data points (400 samples) along with corresponding best beam indices out of 64 beams on Google Map satellite view.}
	\label{fig: Maps}
\end{figure*}

\section{Experimental Results}
    In this section, we describe the results obtained from a series of experiments, followed by datasets description and implementation settings.
	
\subsection{Datasets Description}
    In this work, we consider the DeepSense6G dataset \cite{alkhateeb2023deepsense}, a collection of real-world mmWave sensing and communications measurements, to evaluate the effectiveness of our proposed top-$M$ beam prediction solution presented in Section III. Specifically, following the consistency of our considered system model and formulated problem, we adopt scenario 36 (24,800 samples), scenario 37 (31,000 samples), scenario 38 (36,000 samples), and 39 (20,400 samples). These scenarios are captured with a testbed setup including two moving vehicles (namely, unit $1$ and unit $2$) deployed in diverse outdoor inter-city and urban scenarios having long and shorter distances, respectively, and the data is collected at Tempe, Phoenix, Scottsdale, and Chandler of Arizona during both day as well as night times. 
    
    Here, the unit $1$ acts as a receiver $R_x$, whereas the another unit $2$ acts as a transmitter $T_x$, and both are operating at 60 GHz carrier frequency. Besides, the unit $1$ employs four numbers of mmWave phased arrays, and each phased array has 16 elements ($N_R = 16$) uniform linear arrays (ULA) facing right, left, back, and front directions on the vehicle. Further, the number of pre-defined and over-sampled codebook beams at each phased array of unit $1$ is $64$ ($|\mathbfcal{Q}| = 64$), whereas the unit $2$ has one antenna element, leading to total $64 \times 1$ = $64$ beam pairs.
    
    In particular, the unit $2$ continually performs omni-directional transmission by utilizing its one antenna element, at the same time, the unit $1$ collects the measurements of received power at each beams by making a full beam sweeping. The obtained received power at beams are represented as beam indices ($\mathcal{I} \in \{1, 2, 3 … , 64\}$). Moreover, both the unit $1$ and unit $2$ carry GPS Real Time Kinematics sensors to obtain the real-time vehicle position data, i.e., latitude and longitude values. The data collection maintains a gap between two consecutive samples is $0.1$ second, which makes sampling rate as $10$ samples/second. At each time instant, the unit $1$ records the synchronous data of location coordinates of unit $2$ along with corresponding received powers and optimal beam indices (referred as ground truth indices). As an example, the Figs. \ref{fig: Maps}a, b, c, and d illustrate the visual representation of $400$ location data points of $R_x$ corresponding to individual ground truth beams associated with maximum received powers for all V2V scenarios.

\begin{figure*} [!t]
\begin{subfigure}[b]{0.50\textwidth}
	\centering
	\includegraphics[width=6.5cm]{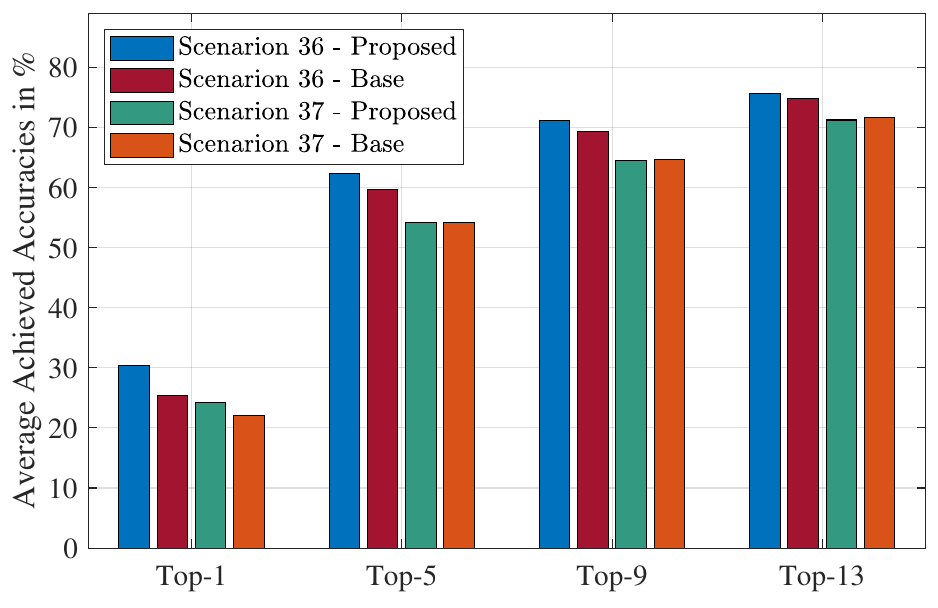} 
	\caption{\centering \scriptsize Accuracies: For scenario 36 and scenario 37}
\end{subfigure}
\begin{subfigure}[b]{0.50\textwidth}
    \centering
    \includegraphics[width=6.5cm]{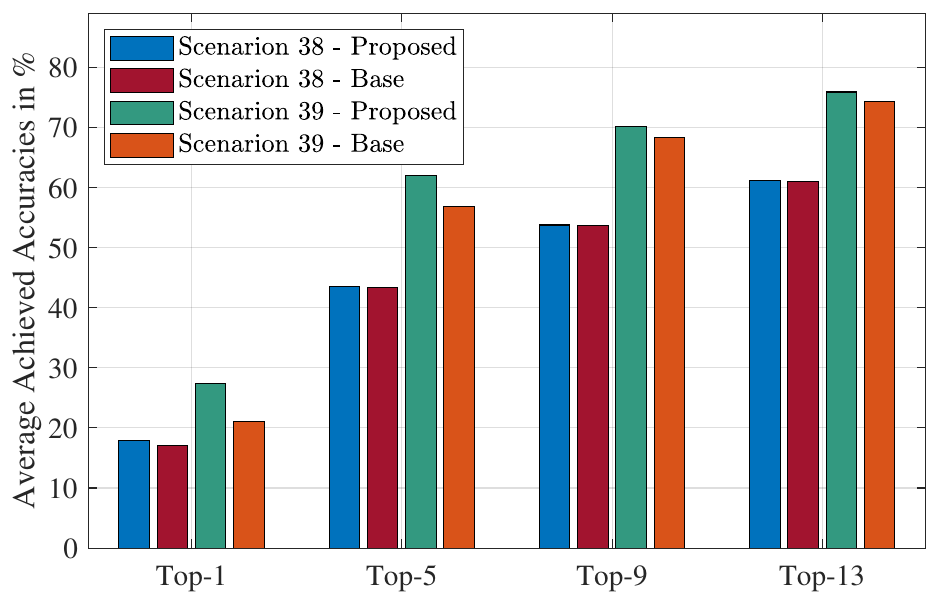}
	\caption{\centering \scriptsize Accuracies: For scenario 38 and scenario 39}
\end{subfigure}

\vspace{2mm}

\begin{subfigure}[b]{0.50\textwidth}
	\centering
	\includegraphics[width=6.5cm]{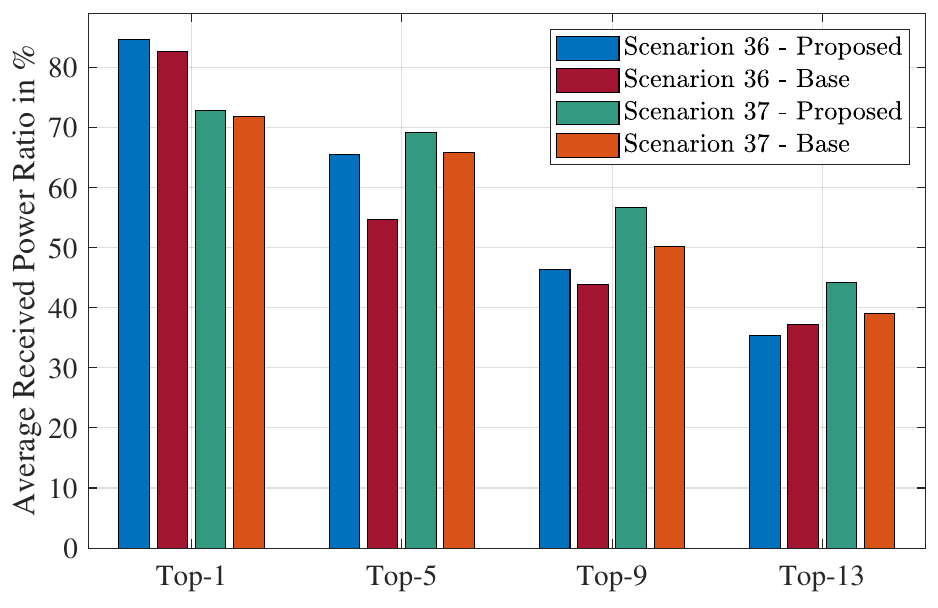}
	\caption{\centering \scriptsize Received power: For scenario 36 and scenario 37}
\end{subfigure}
\begin{subfigure}[b]{0.50\textwidth}
    \centering
    \includegraphics[width=6.5cm]{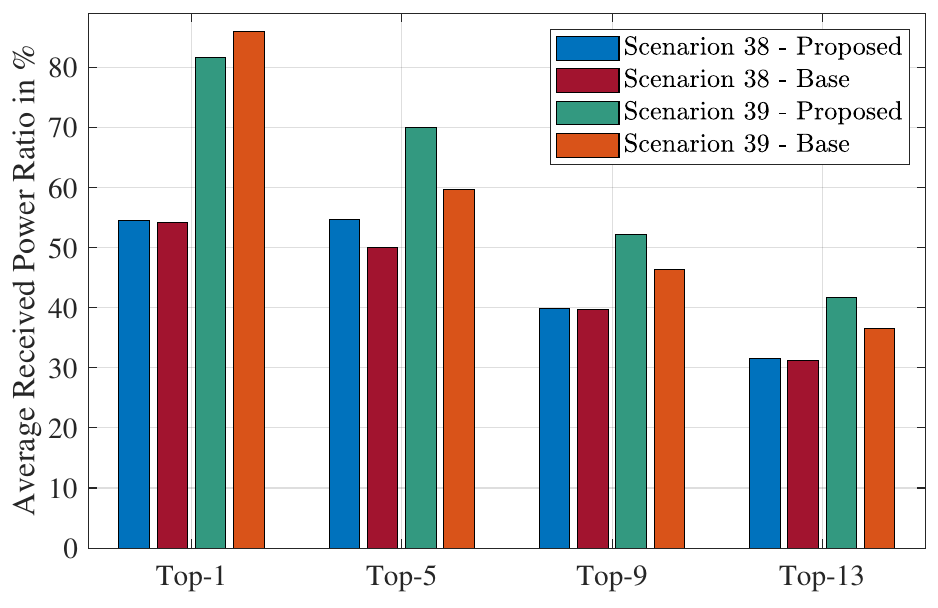}
    \caption{\centering \scriptsize Received power: For scenario 38 and scenario 39}
\end{subfigure}
    \caption{Performance comparison of average achieved accuracies and received power ratio in percentage for all considered vehicle-to-vehicle scenarios.}
	\label{fig: Performances}
\end{figure*}
    
\subsection{Implementation Details}
    We carry out two sets of experiments, for the designed deep learning model as well as a baseline work, on the considered real-world datasets. For the baseline work, we consider the work proposed in \cite{va2017inverse}, and the key reason behind this choosing is because of having same beam search overhead with ours, that is, the overhead is dependent on $M$. 
    
    In particular, the baseline work consists of four steps for beam selection: (i) first, the receiver $R_x$ divides the coverage zone into a number of uniform location bins, and contributing vehicles help to perform channels measurements so that the $R_x$ can build a database including fingerprints of channel characteristics (e.g., received power strengths) and corresponding beams for specific locations bins in advance, (ii) then, the intended transmitter $T_x$ makes a training request and a query along with its position information to the receiver $R_x$ using sub-6GHz bands, (iii) after that, the $R_x$ acknowledges and shares the candidate beams from its database to the $T_x$ to initiate beam training, and (iv) at the end, upon received the acknowledgement, the $T_x$ proceed to beam training following the candidate beams and establish communications over mmWave link after receiving the feedback on the best beam index (among the candidates) from the $R_x$.
    
    \textit{Settings and Parameters:} The experiments are conducted on a computing device having Intel Core i7-10875H CPU with 32 GB RAM and NVIDIA GeForce RTX 2080 Super GPU with 8 GB memory. In particular, the deep learning model is developed by PyTorch and CUDA toolkit 11.7. 
    
    The fixed learning rate and weight decay values of the utilized Adam optimizer are set to $0.01$ and $1 \times 10^{-4}$, respectively, while the batch size $N_b$ is 128, and the model is trained up to total $30$ epochs. We split the data resources of each scenario into $60$\%, $20$\%, and $20$\% for training, validation, and testing, respectively. We utilize this ratio in the proposed solution, whereas we use $80$\% for training and $20$\% for testing in the baseline work. Particularly, we keep the same testing part in both experiments for the purpose of fair comparison.

\subsection{Performance Evaluation}
    For better analysis and illustration of the performance evaluation, we adopt the following two important metrics to measure the performances. Both matrices are averaged over total number of testing samples. 
    
    • \textit{Top-$M$ Accuracy:} The results of number of correct top-$M$ predictions from the model during testing time in relation to the total number of predictions, which can be defined as follows.
    
$$
    Acc_{top-M} = \frac{1}{N_{test}}\sum_{i=0}^{N_{test} - 1}\frac{\mid I^*_t \cap \hat{I}_t \mid}{\mid \hat{I}_t \mid}
$$
    where, $N_{test}$, $I^*_t$, and $\hat{I}_t$ denote the number of test samples, optimal beams at $t^{th}$ time, and top-$M$ beams at $t^{th}$ time, respectively. 

    • \textit{Received Power Ratio:} The ratio between received power in downlink associated with predicted beams and received power from ground-truth beams. Intuitively, this metric in terms of percentage can represent how much percentage of received power can be possibly obtained from the link status. It can be defined as follows.

$$
    R_{\mathcal{P}_t} = \frac{1}{N_{test}}\sum_{i=0}^{N_{test} - 1}\frac{\hat{\mathcal{P}_t}}{\mathcal{P}_t^{(gt)}}
$$
    where, $\hat{\mathcal{P}_t}$ and $\mathcal{P}_t^{(gt)}$ are the received power from predicted beams and ground-truth beams, respectively, at time instant $t$.
    
    First, we compare the performance of this proposed solution with the baseline in terms of accuracy under Top-$M$ beams, in particular, we choose the top beam candidates $M \in \{1, 5, 9, 13\}$. As depicted in Figs. \ref{fig: Performances}a and \ref{fig: Performances}b, we present the results for four scenarios, and it can be observed that our proposed solution outperforms the baseline almost at every top-$M$ beam selection accuracies. For instance, with the same beam search overheads, the top-1 prediction in scenario 36 is performed nearly $19.67$\% increased average accuracy than the baseline, which is made possible by applying trained deep learning model. For another, we also depict the results of received power ratio performances in Figs. \ref{fig: Performances}c and \ref{fig: Performances}d. For example, the results show that with top-1 predicted beams, $84.58$\% received power ratio can be possibly achieved on average in scenario 36. All average results obtained from the experiments after running $5$ times. While quantifying the end-to-end latency performance has already been explored in other works, our emphasis of this proposed solution lies in accomplishing the maximum received power from the communications links. The implementation codes are publicly available at Github\footnote{\url{https://github.com/mbaqer/V2V-mmWave-Beamforming}}.
	
\section{Conclusion}
    In this paper, we have presented a proposed beam selection solution to improve the achieved downlink received power in vehicle-to-vehicle connectivity enabled by 60 GHz mmWave communications. In this solution, we have utilized deep learning model to learn from the out-of-band vehicular position information and predict the top-$M$ beams (a subset of beams). This predicted beams has also helped to reduce the beam searching space, thereby addressing the beam searching overheads limitation of mmWave communications. At the end, experiments on real-world datasets have shown the effectiveness and applicability of the proposed solution, which has indicated the benefits of exploring the side-information in mmWave beamforming. As a future work, this work can be potentially extended by including multiple sensing information together and performing the model fusion.  

    \textbf{Acknowledgment:} This research is supported by National Science Foundation (NSF) under the grant number \# 2010366.


\ifCLASSOPTIONcaptionsoff
  \newpage
\fi

\bibliographystyle{IEEEtran}
\bibliography{bibliography.bib}

\end{document}